# Diffusion leaky LMS algorithm: analysis and implementation

Lu Lu, Haiquan Zhao*

*School of Electrical Engineering, Southwest Jiaotong University, Chengdu, China.*

**ABSTRACT**—The diffusion least-mean square (dLMS) algorithms have attracted much attention owing to its robustness for distributed estimation problems. However, the performance of such filters may change when they are implemented for suppressing noises from speech signals. To overcome this problem, a diffusion leaky dLMS algorithm is proposed in this work, which is characterized by its numerical stability and small misadjustment for noisy speech signals when the unknown system is a lowpass filter. Finally, two implementations of the leaky dLMS are introduced. It is demonstrated that the leaky dLMS can be effectively introduced into a noise reduction network for speech signals.

**Keywords***:* Distributed estimation; Diffusion LMS; Leaky LMS; Noisy speech signal.

## 1. Introduction

Since the simple structure and low computational burden, the least-mean square (LMS) algorithm has become a widely used adaptive filter. However, in practice, it is well known that direct implementation of the conventional LMS algorithm can be problematic. Such problems can be outlined as follows. (i) Numerical problem, caused by inadequacy of excitation in the input signal; (ii) stagnation behaviour, often occur in low input signal scenarios, making the performance degradation of the algorithm. To solve these problems, the leaky LMS algorithm [1] as its variants were proposed in diverse fields, such as active noise control (ANC) [2], channel estimation [3], and nonlinear acoustic echo cancellation (NLAEC) [4].

Recently, to estimate some parameters of interest from the data collected at nodes distributed over a geographic region, several distributed estimation algorithms were developed, including incremental [5-6], and diffusion algorithms [7-10], etc. In the incremental strategy, a cyclic path is required the definition over the nodes, and this technique is also sensitive to link failures [11]. On the other hand, the diffusion method, is widely used because of its ease of implementation. In this strategy, each node

*E-mail addresses*: lulu@my.swjtu.edu.cn (L. Lu), hqzhao@home.swjtu.edu.cn (H. Zhao)

communicates with a subset of its neighbours, which achieves more data from their neighbours with moderate amount of communications. Particularly, in [12], two versions of the diffusion LMS (dLMS) algorithms, adapt-then-combine (ATC) and combine-then-adapt (CTA) were proposed, based on the different orders of *adaptation* and *combination* steps. Note that the ATC method outperforms the CTA method in all cases, and better performance can be achieved if the measurement are shared. Due to its merit, the ATC version of the dLMS algorithm (ATC dLMS) has been introduced to subband adaptive filter [13], information theoretic learning (ITL) [14] to obtain improved performance.

Motivated by these considerations, in this paper, we proposed a leaky dLMS algorithm with two combination strategies-ATC and CTA, resulting in ATC leaky dLMS and CTA leaky dLMS algorithms. Very recently, a distributed incremental leaky LMS was proposed to surmount the drift problem of dLMS algorithm [15]. Unfortunately, it is derived from the incremental approach, which prohibits its practical applications. Compared with the existing algorithms, the leaky dLMS algorithm is derived by minimizing the instantaneous leaky objective function rather than the mean square error cost function. Moreover, it has a superiority performance via ATC strategy.

*Notation*: Throughout this paper, we use $\{\cdot\}$ to represent a set, $|\cdot|$ denotes absolute value of a scalar, $(\cdot)^T$ denotes transposition, $E\{\cdot\}$ denotes expectation, $tr\{\cdot\}$ denotes trace operator, $\|\cdot\|_2^2$ denotes $l_2$-norm, and $\{\cdot\}^o$ is the real value of parameter. Besides, we use boldface and normal letters to denote the random quantities and deterministic quantities, respectively.

## 2. Diffusion LMS strategies

Consider a network of $N$ sensor nodes distributed over a geographic area. At each time instant $i$, each sensor node $k \in \{1, 2, ..., N\}$ has access to the realization $\{d_k(i), u_{k,i}\}$ of some zero-mean random process $\{\mathbf{d}_k(i), \mathbf{u}_{k,i}\}$, and $\mathbf{d}_k(i)$ is a scalar and $\mathbf{u}_{k,i}$ is an regression vector with length $M$. Suppose these measurements follow a standard model given by:

$$\mathbf{d}_k(i) = \mathbf{u}_{k,i} w^o + \mathbf{v}_k(i) \tag{1}$$

where $w^o \in \mathbb{C}^{M\times 1}$ is the unknown parameter vector, and $\mathbf{v}_k(i)$ is the measurement noise with variance $\sigma_{v,k}^2$. Here, we assume that $\mathbf{u}_{k,i}$ and $\mathbf{v}_k(i)$ are spatially independent and independent identically distributed (i.i.d.), and $\mathbf{v}_k(i)$ is independent of $\mathbf{u}_{k,i}$.

The dLMS algorithm is obtain by minimizing a linear combination of the local mean square error (MSE):

$$J_k^{loc}(w) = \sum_{l\in N_k} a_{l,k} E|\mathbf{e}_{l,i}|^2 = \sum_{l\in N_k} a_{l,k} E|\mathbf{d}_l(i) - \mathbf{u}_{l,i} w|^2 \tag{2}$$

where $N_k$ is the set of nodes with which node $k$ shares information (including $k$ itself). The weighting coefficient $\{a_{l,k}\}$ are real, non-negative, and satisfy: $\sum_{l=1}^{N} a_{l,k} = 1$.

The dLMS algorithm obtains the estimation via two steps, *adaptation* and *combination*. According to the order of these two steps, the diffusion LMS algorithm is classified into the ATC dLMS and CTA dLMS algorithms. The updation equation of ATC dLMS can be expressed as

$$\begin{cases} \varphi_{k,i} = w_{k,i-1} + \mu \sum_{l \in N_k} c_{l,k} u_{l,i}^T (d_l(i) - u_{l,i} w_{k,i-1}) & (adaptation) \\ w_{k,i} = \sum_{l \in N_k} a_{l,k} \varphi_{l,i} & (combination) \end{cases} \tag{3}$$

where $\mu$ is the step size (learning rate), and $\varphi_k$ is the local estimates at node *k*. The weighting coefficients $\{c_{l,k}\}$ is the real, non-negative, satisfying the condition $c_{l,k} = a_{l,k} = 0$ if $l \notin N_k$. Similarly, the CTA dLMS algorithm can be given as

$$\begin{cases} \varphi_{k,i-1} = \sum_{l \in N_k} a_{l,k} w_{l,i-1} & (combination) \\ w_{k,i} = \varphi_{k,i-1} + \mu \sum_{l \in N_k} c_{l,k} u_{l,i}^T (d_l(i) - u_{l,i} \varphi_{k,i-1}) & (adaptation). \end{cases} \tag{4}$$

# 3. Proposed diffusion leaky LMS algorithm

In this section, the leaky dLMS is proposed to address the two problems mentioned above. For each node *k*, we seek an estimate of $w^o$ by minimizing the following cost function:

$$\begin{aligned} J_k^{loc}(w) &= \sum_{l \in N_k} a_{l,k} E |\mathbf{e}_{l,i}|^2 + \gamma w^T w \\ &= \sum_{l \in N_k} a_{l,k} E |\mathbf{d}_l(i) - \mathbf{u}_{l,i} w|^2 + \gamma w^T w \end{aligned} \tag{5}$$

where γ>0 is the leakage coefficient, and *w* is the estimate of $w^o$. Using the steepest descent algorithm, yields

$$\nabla_w J_k^{loc}(w) = \sum_{l \in N_k} a_{l,k} \frac{\partial \{E |\mathbf{e}_{l,i}|^2 + \gamma w^T w\}}{\partial w}. \tag{6}$$

Therefore, the updating of proposed algorithm for estimating $w^o$ at node *k* can be derived from the steepest descent recursion

$$w_{k,i} = w_{k,i-1} - \mu \sum_{l \in N_k} a_{l,k} \frac{\partial \{E |\mathbf{e}_{l,i}|^2 + \gamma w^T w\}}{\partial w} |_{w_{k,i-1}}. \tag{7}$$

Under the linear combination assumption [12,14], let us define the linear combination $w_{k,i}$ at node *k* as

$$w_{k,i-1} = \sum_{l \in N_k} a_{l,k} \varphi_{l,i-1}. \tag{8}$$

Introducing (8) to (7), we can obtain an iterative formula for the intermediate estimate

$$\begin{aligned}\varphi_{l,i} &= \varphi_{l,i-1} - \mu \frac{\partial E|\mathbf{e}_{l,i}|^2 + \gamma w^T w}{\partial w}\Big|_{w_{k,i-1}} \\ &\approx \varphi_{l,i-1} - \mu \frac{\partial E|\mathbf{e}_{l,i}|^2 + \gamma w^T w}{\partial w}\Big|_{w_{l,i-1}} \\ &= (1-\gamma\mu)\varphi_{l,i-1} + \mu u_{l,i}^T e_{l,i}.\end{aligned} \tag{9}$$

Hence, we obtain the leaky dLMS algorithm by transforming the steepest-descent type iteration of (7) into a two-step iteration

$$\begin{cases}\varphi_{k,i} = w_{k,i-1} - \mu \dfrac{\partial E|\mathbf{e}_{k,i}|^2 + \gamma w^T w}{\partial w}\Big|_{w_{k,i-1}} \\ w_{k,i} = \sum_{l\in N_k} a_{l,k}\varphi_{l,i}.\end{cases} \tag{10}$$

In *adaptation* step of (10), the local estimate $\varphi_{k,i-1}$ is replaced by linear combination $w_{k,i-1}$. Such substitution is reasonable, because the linear combination contains more data information from neighbor nodes than $\varphi_{k,i-1}$ [12]. Then, we extend the leaky dLMS algorithm to its ATC and CTA forms.

•**ATC leaky dLMS algorithm**:

$$\begin{cases}\varphi_{k,i} = (1-\mu\gamma)w_{k,i-1} + \mu\sum_{l\in N_k} c_{l,k}u_{l,i}^T(d_l(i) - u_{l,i}w_{k,i-1}) & (adaptation) \\ w_{k,i} = \sum_{l\in N_k} a_{l,k}\varphi_{l,i} & (combination).\end{cases} \tag{11}$$

•**CTA leaky dLMS algorithm**:

$$\begin{cases}\varphi_{k,i-1} = \sum_{l\in N_k} a_{l,k}w_{l,i-1} & (combination) \\ w_{k,i} = (1-\mu\gamma)\varphi_{k,i-1} + \mu\sum_{l\in N_k} c_{l,k}u_{l,i}^T(d_l(i) - u_{l,i}\varphi_{k,i-1}) & (adaptation).\end{cases} \tag{12}$$

## 4. Simulation Results

In the following, we evaluate the performance of the proposed algorithm for noise reduction network with 20 nodes. The topology of the network is shown in Fig. 1. The unknown vector of interest is a lowpass filter of order $M$=5, whose coefficients and normalized frequency response are shown in Fig. 2(a-b). The regressors are zero-mean white Gaussian distributed with covariance matrices $R_{u,k} = \sigma_{uk}^2 I$, with $\sigma_{uk}^2$ shown in Fig. 2(c). The performance of the different algorithms is measured in terms of the network mean square deviation (MSD) $\mathrm{MSD} = 10\log_{10}(\sum_{l=1}^{N} \| w_l - w^o \| / N)$. All network MSD curves are obtained by ensemble averaging over 50 independent trials.

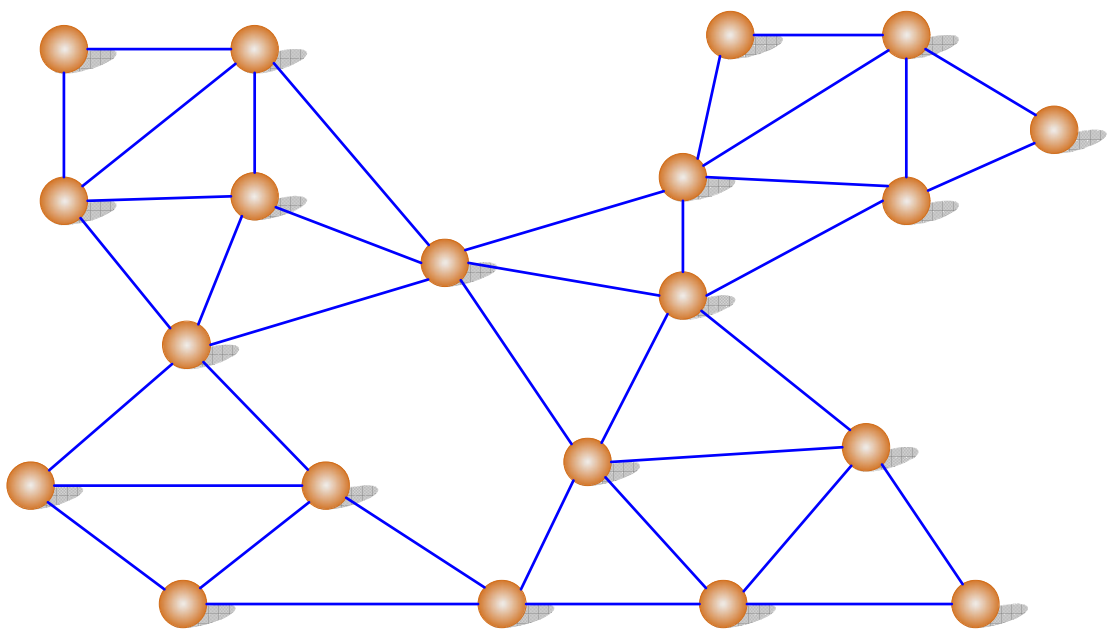

**Fig. 1.** Network topology.

## 5.1 Example 1: Gaussian input

In this example, the Gaussian signal with zero mean and unit variance is employed as the input signal, $a_{l,k}=1/n_k$ is used for $\{a_{l,k}\}$, and $c_{l,k}=1/n_k$ (uniform, [16]), where $n_k$ is the degree of node *k*. The signal-to-noise ratio (SNR) of each node is SNR=0dB, corresponding to the high background noise power.

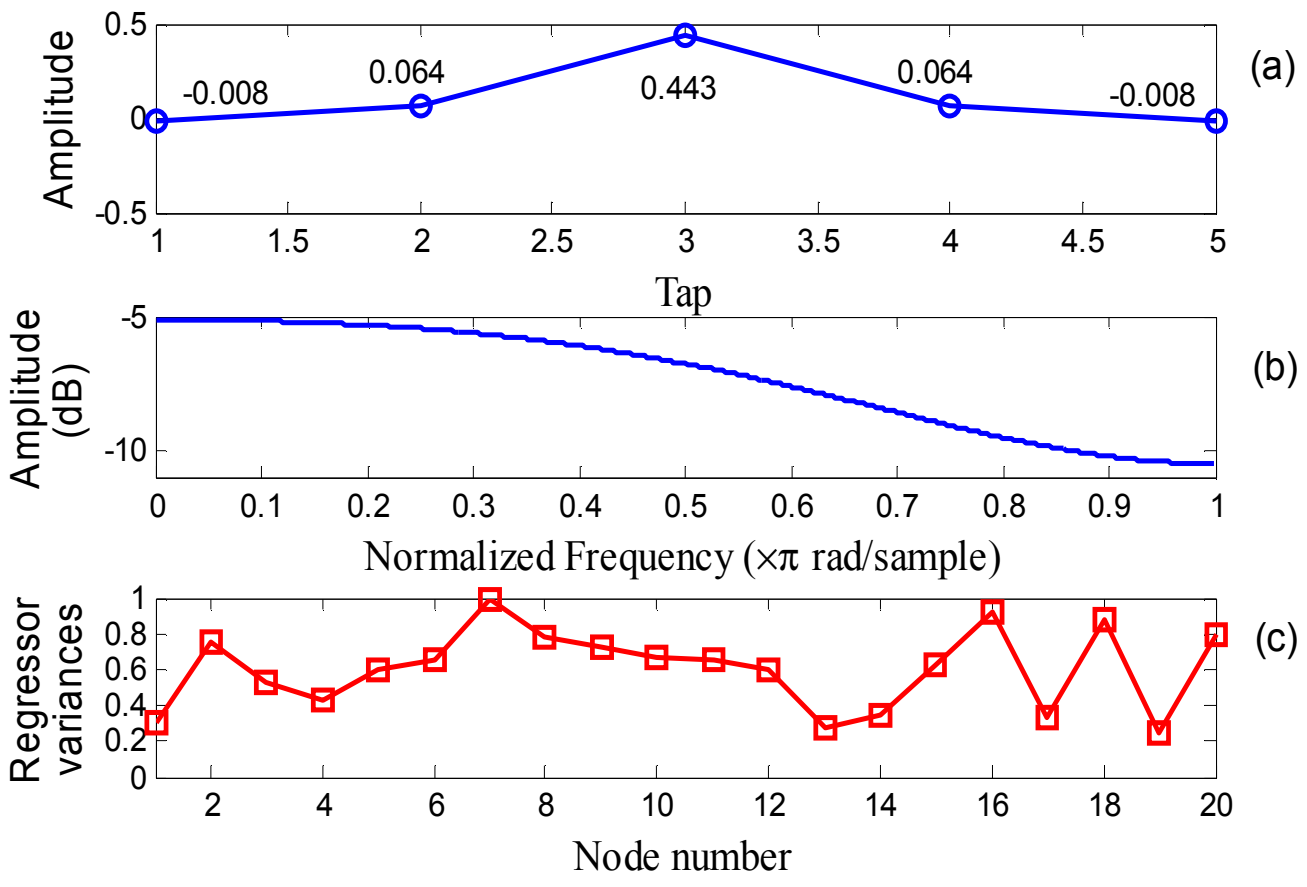

**Fig. 2.** (a) Coefficients of the filter. (b) Frequency responses of the unknown system. (c) Regressor variances.

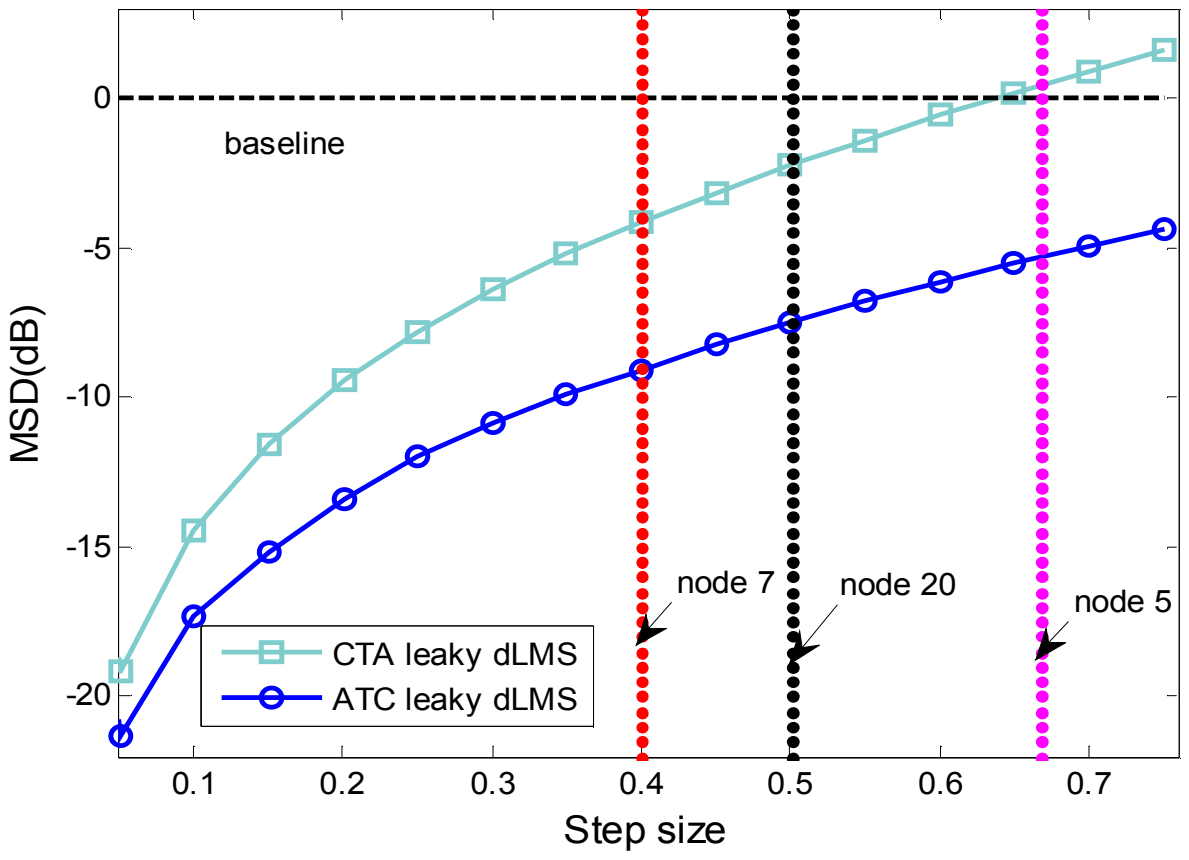

**Fig. 3.** Network MSD of the proposed algorithms versus different step sizes

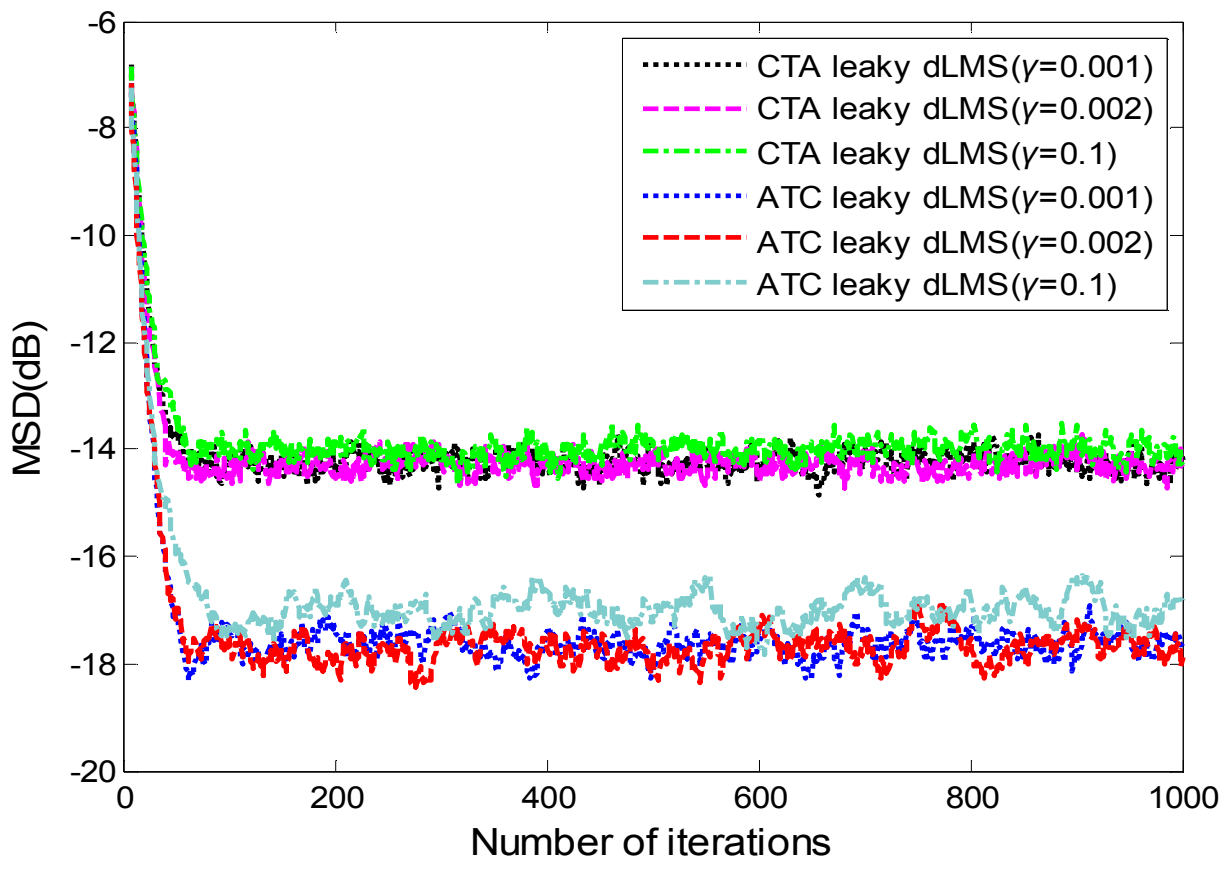

**Fig. 4.** Network MSD of the proposed algorithms versus different $\gamma$ ($\mu$=0.08).

Fig. 3 shows the steady-state network MSD of the leaky dLMS with different step-size. As one can see, the steady-state MSDs of the proposed algorithms increase as the step sizes increase. For $\mu$>0.668 (node 5), the CTA leaky dLMS fails to work. Besides, by using (31), we can obtain the theory upper bounds of step sizes from different nodes. The calculating data indicate that, the upper bound of step size from node 1 is 1.1 (maximum upper bound among 20 nodes), and upper bound of step size from node 7 is about 0.4 (minimum upper bound among 20 nodes). To guarantee all the nodes work well, $\mu$ must be $\mu$<0.4 to ensure that the stability. As a result, we select $\mu$=0.08 in the following simulations. Then, we investigate the performance of the algorithm in different $\gamma$, as shown in Fig. 4. It can be easily observed that the CTA/ATC leaky dLMS algorithm is not sensitive to this choice, but it turns out that the best option is $\gamma$=0.002. Fig. 5 displays a comparison of MSE from the proposed algorithm and existing algorithms. One can see that the proposed algorithm outperforms the conventional dLMS algorithms in terms of steady-state error. After the about iteration 50, all the algorithm reach steady state. In addition, it is obviously shown from these curves that the proposed algorithms reach low misadjustment (−14dB and −18dB) in about iteration 60, whereas the CAT dLMS and ATC dLMS algorithms finally settle at a network MSD value of −12dB and −16dB. All the ATC-based algorithms achieve improved performance as compared with CTA-based algorithms, with the similar initial convergence rate.

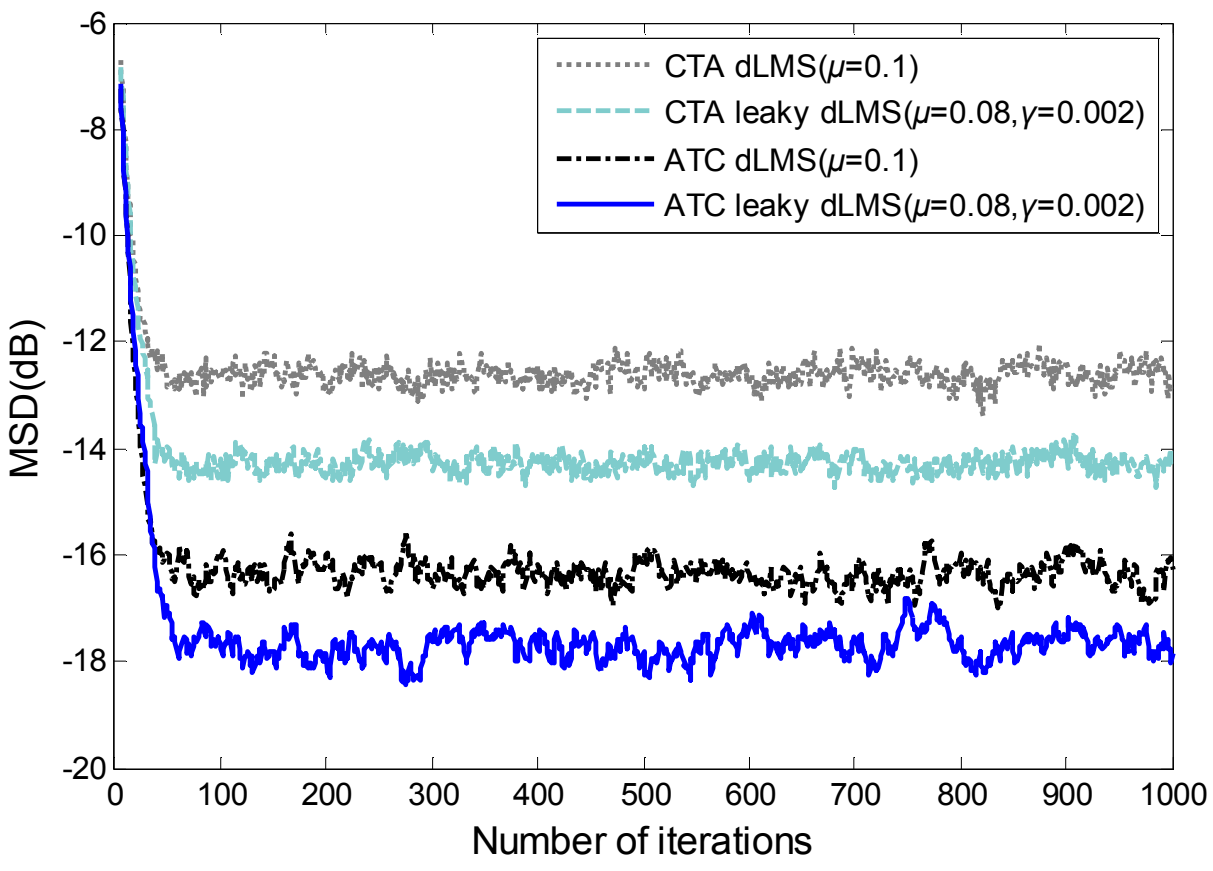

**Fig. 5.** Network MSD learning curves for Gaussian input.

## 5.2 Example 2: Speech input signal

In the second example, the input signal is a speech input (See Fig. 6). In piratical application, different nodes have different speech signals. Therefore, we consider a input signal generated as $\mathbf{u}_{u,k} = \sigma_{uk}^2 \bullet speech\ input$ . The environment and parameters of all the algorithms are same as those of example 1.

In Fig. 7, it demonstrates a comparison of network MSD performance of different algorithms. In this cases, superior performance of the proposed algorithms are obtained because that the leaky method can stabilize the system [1]. The proposed algorithm, which adopts the ATC and CTA approaches, reduces the misadjustment when the speech is selected as the input signal. Meanwhile, the fast convergence rate is obtained. Furthermore, it can be obviously observed from this figure that the misadjustment for the ATC version is less than that of CTA version. To further demonstrate the performance of proposed algorithm, Fig. 8 depicts the waveforms of the simulation result at node 14 ( $\sigma_{uk}^2 = 0.35$ ). The ATC leaky dLMS algorithm, inherits the advantages of the leaky algorithm, and achieves a stability performance in the presence of strong disturbances.

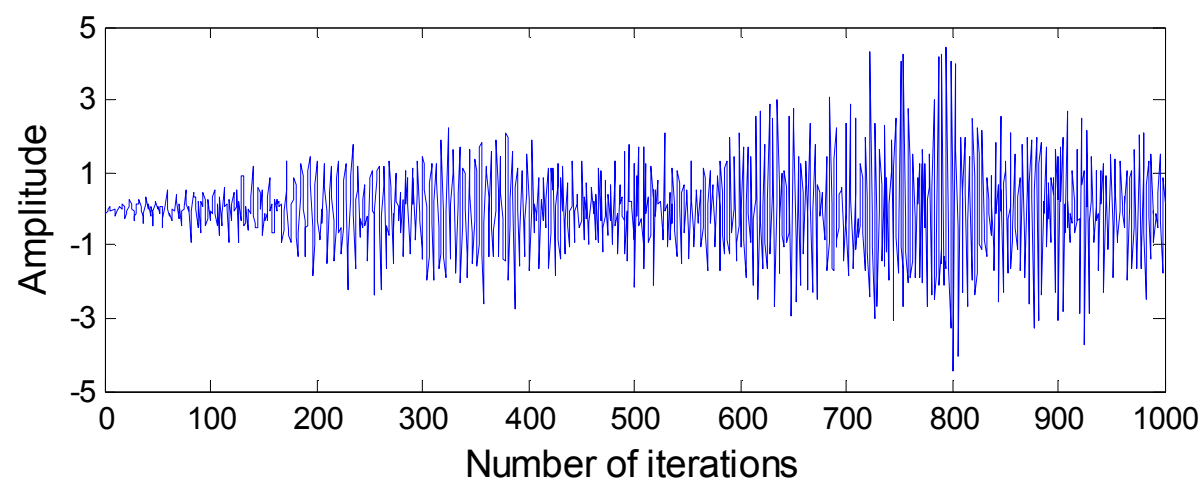

**Fig. 6.** Input speech signals

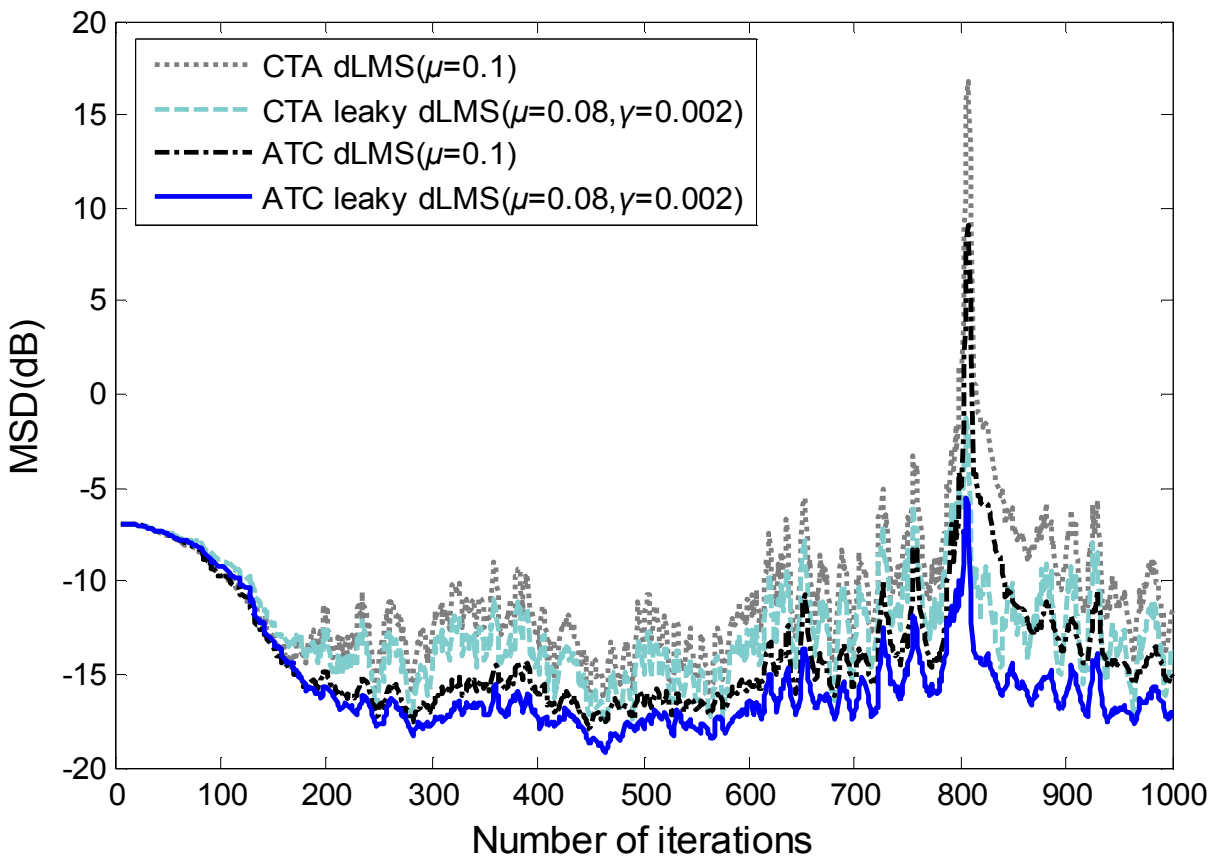

**Fig. 7.** Network MSD learning curves for speech input.

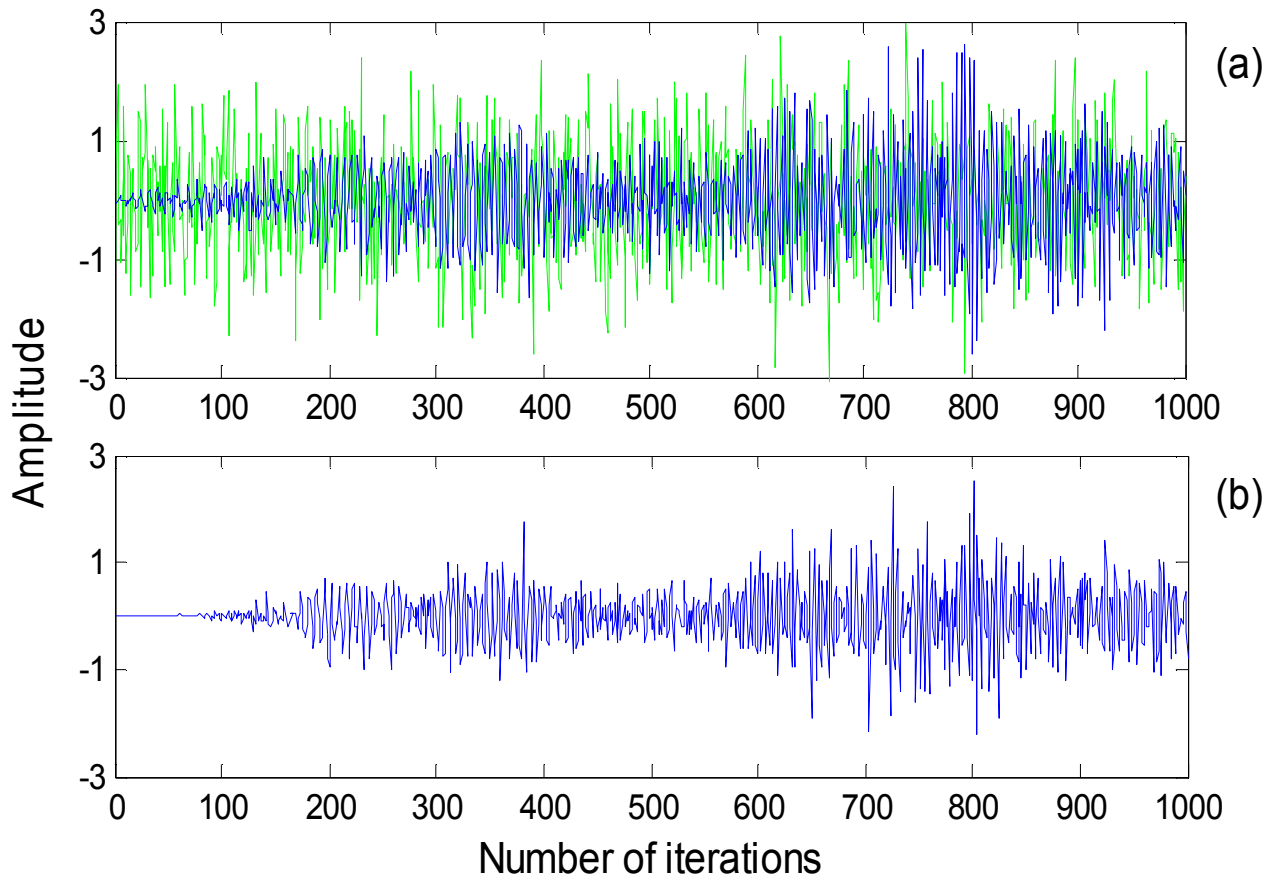

**Fig. 8.** Waveforms of the simulation result at node 14. (a) Noisy speech (green line: noise signal with SNR=0dB; blue line: clean speech). (b) Filtered speech with the ATC leaky dLMS algorithm.

**Table 1** Computation time

| **Algorithms** | **Computation Time(s)** |
|---|---|
| ATC dLMS | 0.6962 |
| **ATC leaky dLMS** | **1.0026** |
| CTA dLMS | 1.1945 |
| **CTA leaky dLMS** | **1.5701** |

To evaluate the computational burden, we have measured the average run execution time of the algorithm on a 2.1-GHz AMD processor with 8GB of RAM, running Matlab R2013a on Windows 7. The computation time of the algorithms is outlined in Table. 1. One can see that the ATC-dLMS algorithm is the fastest one among these distributed algorithms. The proposed algorithm slightly increases the execution time, but the final performance is lower than other algorithms. The CTA algorithms achieve the slower than the ATC algorithms, still with an affordable computation time.

# 5. Conclusion

In this paper, a leaky dLMS algorithm with ATC and CTA strategies has been proposed for distributed estimation. The proposed algorithm updates the local estimation based on leaky method, which can not only maintain the stability for distributed system, but also enhance the performance for noise suppression. Simulation results in the context of noise-reduction system show that the proposed algorithm achieves an improved performance as compared with existing algorithms.

## Acknowledgement

This work was partially supported by National Science Foundation of P.R. China (Grant nos: 61271340, 61571374, 61433011). The first author would also like to acknowledge the China Scholarship Council (CSC) for providing him with financial support to study abroad (No. 201607000050).